\documentstyle[12pt]{article}

\font\goth=eufm10 scaled 1400
\font\Goth=eufm10 scaled 2000
\font\math=msbm10

\def\g{\gamma}

\def\s{\sigma}

\def\S1{\hbox{\rm S$^1$}}
\def\Vect{\hbox{\rm Vect}}

\def\pds#1,#2{\langle #1\mid #2\rangle} %%% PRODUIT SCALAIRE
\def\f#1,#2,#3{#1\colon#2\to#3} %%%  F:A->B

\def\hfl#1{{\buildrel{#1}\over{\hbox to
12mm{\rightarrowfill}}}}

\chardef\s=110
\chardef\g=103

\begin{document}

\title{Extension of the Virasoro and Neveu-Schwartz algebras and
generalized Sturm-Liouville operators.}

\author{Patrick MARCEL, Valentin OVSIENKO\\
{\small C.N.R.S., C.P.T.}\\
{\small  Luminy-Case 907}\\
{\small  F-13288 Marseille Cedex 9, France}\\
\\
Claude ROGER\\
{\small Institut
Girard Desargues, URA CNRS 746}
\\ {\small Universit\'e Claude Bernard - Lyon I}\\
{\small 43
bd. du 11 Novembre 1918}\\
{\small 69622 Villeurbanne Cedex, France}}

\date{}

\maketitle

{\abstract{We consider the universal central extension of the
Lie algebra $\Vect (S^1)${\math \s}$ C^{\infty}(S^1)$.
The coadjoint representation of this Lie algebra has
a natural geometric interpretation by matrix analogues of
the Sturm-Liouville operators. This approach leads to new
Lie superalgebras generalizing the well-known Neveu-Schwartz algebra.}}

\thispagestyle{empty}

\vfill\eject

\section{Introduction}

{\bf 1.1 Sturm-Liouville operators and the action of $\Vect (S^1)$}. 
Let us recall some well-known definitions (cf. e.g. \cite{Car},\cite{Tur}).

Consider the Sturm-Liouville operator:
\begin{equation}
 L=-2c\frac{d^2}{dx^2} +u(x)
\label{s-l}							
\end{equation}
where $c\in {\bf R}$ and $u$ is a periodic potential: $u(x+2\pi)=u(x)\in C^{\infty}(\bf R)$.

\vskip 0,3cm

Let $\Vect (S^1)$ be the Lie algebra of smooth vector field on $S^1$:
$$
f=f(x)\frac{d}{dx},
$$ 
where $f(x+2\pi)=f(x)$, with the commutator
$$
[f(x)\frac{d}{dx},g(x)\frac{d}{dx}]=(f(x)g'(x)-f'(x)g(x))\frac{d}{dx}.
$$

We define a $\Vect (S^1)$-action on the space of Sturm-Liouville operators.

\vskip 0,3cm

Consider a {\it 1-parameter family} of $\Vect (S^1)$-actions on the space of
smooth functions $C^{\infty}(S^1)$:
\begin{equation}
L_{f(x)\frac{d}{dx}}^{(\lambda)}\;a(x)=
f(x)a'(x)-{\lambda}f'(x)a(x).
\label{Lied}
\end{equation}

{\bf Notation}.
1) The operator
$$
L_{f(x)\frac{d}{dx}}^{(\lambda)}=
f(x)\frac{d}{dx}-{\lambda}f'(x)
$$
is called the {\it Lie derivative}.

2) Denote ${\cal F}_\lambda $ the $\Vect (S^1)$-module structure
(\ref{Lied}) on $C^{\infty}(S^1)$.

\vskip 0,3cm

\proclaim Definition. The $\Vect (S^1)$-action on $L$ is defined
by the commutator with the Lie derivative:
\begin{equation}
[L_{f\frac{d}{dx}},L]:=
L_{f\frac{d}{dx}}^{(-\frac{3}{2})}\circ L-L\circ L_{f\frac{d}{dx}}^{(\frac{1}{2})}
\label{c-r}
\end {equation}\par

The result of this action is a {\it scalar operator}, i.e. the operator of multiplication 
by a function:
\begin{equation} 
[L_{f(x)\frac{d}{dx}},L]=f(x)u'(x)+2f'(x)u(x)-cf'''(x)
\label{act}
\end {equation}

{\bf Remark}. The argument $a$ of the operator (\ref{Lied}) 
has a natural geometric interpretation as a 
{\it tensor-densities} on $S^1$  of
degree $-\lambda $:
$$
a=a(x)(dx)^{-\lambda}.
$$
One obtains a natural realization of the Sturm-Liouville operator
as an operator on tensor-densities:
$$
  L:{\cal F}_{\frac{1}{2}}\to{\cal F}_{-\frac{3}{2}}
$$
 (cf. \cite{Tur}).

\vskip 0,3cm

{\bf 1.2 The coadjoint representation of the Virasoro algebra}. 
The {\it Virasoro algebra} is the unique (up to isomorphism) non-trivial central
extension of $\Vect (S^1)$. It is given by the Gelfand-Fuchs cocycle:
\begin{equation}
 c(f(x)\frac{d}{dx},g(x)\frac{d}{dx})=\int_{0}^{2\pi}f'(x)g''(x)dx
\label{Gel}
\end {equation}
The Virasoro algebra is therefore a Lie algebra on the space
$\Vect (S^1)\oplus\bf R$ with the commutator
$$
[(f,\alpha),(g,\beta)]=([f,g]_{\Vect(S^1)},c(f,g)).
$$

A deep remark of A.A Kirillov and G.Segal (see \cite{Kir1},\cite{Seg}) is:
{\it the $\Vect (S^1)$-action (\ref{act}) coincides 
with the coadjoint action of the Virasoro algebra}.

\vskip 0,3cm

Let us give the precise definitions.

Consider the space $C^{\infty}(S^1)\oplus\bf R$ and a pairing between this space and
the Virasoro algebra:
$$
\langle(u(x),c),\;(f(x)\frac{d}{dx},\alpha)\rangle=\int_{0}^{2\pi}u(x)f(x)dx\;+\;
c\alpha. 
$$
Space $C^{\infty}(S^1)\oplus\bf R$ is identified with a part
of the dual space to the Virasoro algebra. It is called the {\it regular part}
(see \cite{Kir1}).

\proclaim Definition. The coadjoint action 
of the Virasoro algebra on $C^{\infty}(S^1)\oplus\bf R$ is 
defined by:
$$
\langle ad^*_{(f\frac{d}{dx},\alpha)}(u(x),\;c),\;(g\frac{d}{dx},\;\beta)\rangle:=
-\langle (u(x),\;c),\;[(f\frac{d}{dx},\;\alpha),\;(g\frac{d}{dx},\;\beta)]\rangle
$$\par

It is easy to calculate the explicit formula.
The result 
is as follows:
$$
ad^*_{(f(x)\frac{d}{dx},\;\alpha)}(u(x),\;c)=
((L_{f(x)\frac{d}{dx}}^{(-2)}\;u(x)-cf'''(x),\;0),
$$
where $L^{(2)}_f$ is the operator of Lie derivative (\ref{Lied}).
This action coincides with the $\Vect(S^1)$-action (\ref{act})
on the space of Sturm-Liouville operators.

\vskip 0,3cm

{\bf Remark}. 1) Note, that the coadjoint action of the Virasoro algebra is in fact,
a $\Vect (S^1)$-action (the center acts trivialy).

2) The regular part of the dual space to the Viresoro algebra
is can be interpreted as a deformation of the $\Vect (S^1)$-module ${\cal F}_{-2}$.

\section{Central extension of $Vect (S^1)${\math \s}$ C^{\infty}(S^1)$}

Consider the semi-direct product:
${\cal G}=Vect (S^1)${\math \s}$ C^{\infty}(S^1)$.
This Lie algebra has a 3-dimensional central extension
given by the non-trivial $2$-cocycles: 
\begin{equation}
\matrix{
\sigma_1((f\frac{d}{dx},\;a),\;(g\frac{d}{dx},\;b))=
\displaystyle\int_{S^1}f'(x)g''(x)dx\hfill\cr\noalign{\bigskip}
\sigma_2((f\frac{d}{dx},\;a),\;(g\frac{d}{dx},\;b))=
\displaystyle\int_{S^1}(f''(x)b(x)-g''(x)a(x))dx\hfill\cr\noalign{\bigskip}
\sigma_3((f\frac{d}{dx},\;a),\;(g\frac{d}{dx},\;b))=
\displaystyle2\int_{S^1}a(x)b'(x)dx\hfill\cr\noalign{\bigskip}
}
\label{coc}
\end{equation}
Let us denote {\goth \g} the Lie algebra defined by this extension.

As a vector space  {\goth \g}$=Vect (S^1)${\math \s}$ C^{\infty}(S^1)\oplus{\bf R}^3$.
The commutator in {\goth \g} is:
\begin{equation}
[(f\frac{d}{dx},a,{\bf \alpha}),\;
(g\frac{d}{dx},b,{\bf \beta})]=
( (fg'-f'g)\frac{d}{dx} ,\;fb'-ga',\; {\bf \sigma})
\label{[]}
\end{equation}
where ${\bf \alpha}=(\alpha_1,\alpha_2,\alpha_3),
{\beta}=(\beta_1,\beta_2,\beta_3)\in {\bf R}^3$ and
$\sigma=(\sigma_1, \sigma_2, \sigma_3)$ are
the 2-cocycles given by the formula (\ref{coc}).

\vskip 0,3cm

The Lie algebra {\goth \g} is known in physical literature 
(see \cite{Arb},\cite{Har}).
It was shown in \cite{OR} that the cocycles (\ref{coc})
define the {\it universal} central extension\footnote{it
makes sense since $H_1(Vect (S^1)${\math \s}$ C^{\infty}(S^1))=0$.}
the Lie algebra $Vect (S^1)${\math \s}$ C^{\infty}(S^1)$.
This means, $H^2(Vect (S^1)${\math \s}$ C^{\infty}(S^1))={\bf R}^3$.

\vskip 0,3cm

In this paper we define a space of matrix linear differential operators
generalizing the Sturm-Liouville operators. This space 
give a natural geometric realization of the coadjoint representation
of the Lie algebra {\goth \g}.
We hope that such a realization can be usefull for the theory of 
KdV-type integrable systems related to the Lie algebra {\goth \g}
as well as for studying of the coadjoint orbits of {\goth \g}
(cf. \cite{Kir1} for the Virasoro case).
Remark here that interesting results concernig coadjoint orbits of
{\goth \g} has been recently obtained in \cite{Kap}.

\section{Matrix Sturm-Liouville operators}
{\bf Definition}.
Consider the following
matrix linear differential operators
on $C^{\infty}(S^1)\oplus C^{\infty}(S^1)$:
\begin{equation}
{\cal L}=\left(
\matrix{
-2c_1\displaystyle\frac{d^2}{dx^2} +u(x) & 2c_2\displaystyle\frac{d}{dx}+v(x) \hfill\cr\noalign{\smallskip}
-2c_2\displaystyle\frac{d}{dx}+v(x) & 4c_3\hfill\cr\noalign{\smallskip}
}
\right)
\label{ope}
\end{equation}
where $c_1,c_2,c_3\in {\bf R}$ and $u=u(x),v=v(x)$
are $2\pi$-periodic functions.

The $Vect (S^1)$-action on the space of operators (\ref{ope}) is defined,
as in the case of Sturm-Liouville operators (\ref{s-l}),
by commutator with the Lie derivative.
We consider 
 ${\cal L}$ as an operator on $Vect (S^1)$-modules:
$$
{\cal L}:{\cal F}_{\frac{1}{2}}\oplus{\cal F}_{-\frac{1}{2}}\to
{\cal F}_{-\frac{3}{2}}\oplus{\cal F}_{-\frac{1}{2}}.
$$
We will show that there exists a reacher
structure on the space of operators (\ref{ope}).
Namely, we will define an action of the semi-direct product
$Vect (S^1)${\math \s}$ C^{\infty}(S^1)$.

\subsection{$Vect (S^1)${\math \s}$ C^{\infty}(S^1)$-module structure}
Let us define a 1-parameter family
of
$Vect (S^1)${\math \s}$ C^{\infty}(S^1)$-modules 
on the space $C^{\infty}(S^1)\oplus C^{\infty}(S^1)$:
\begin{equation}
T^{(\lambda)}_{\displaystyle(f(x)\frac{d}{dx},\;a(x))}
\left(
\matrix{
\phi(x)\hfill\cr\noalign{\smallskip}
\psi(x)\hfill\cr\noalign{\smallskip}
}
\right)=
\left(
\matrix{
L_{f\frac{d}{dx}}^{(\lambda)}\;\phi(x)\hfill\cr\noalign{\smallskip}
L_{f\frac{d}{dx}}^{(\lambda-1)}\;\psi(x) -{\lambda}a'(x){\phi}(x)\hfill\cr\noalign{\smallskip}
}
\right)
\label{act'}
\end {equation}
where ${\phi}(x),{\psi}(x)\in C^{\infty}(S^1)$.
Verify, that this formula defines a $Vect (S^1)${\math \s}$ C^{\infty}(S^1)$-action:
$$
\Bigl[T^{(\lambda)}_{(\displaystyle f\frac{d}{dx},\;a)},
\;T^{(\lambda)}_{(\displaystyle g\frac{d}{dx},\;b)}\Bigr]=
 T^{(\lambda)}_{\displaystyle((fg'-f'g)\frac{d}{dx},\; fb'-ga')}.
$$

\vskip 0,3cm

\proclaim Definition. Define the
$Vect (S^1)${\math \s}$ C^{\infty}(S^1)$-action 
 on the space of the operators (\ref{ope}) by:
\begin{equation} 
\Bigl[T_{\displaystyle(f\frac{d}{dx},a)},{\cal L}\Bigr]:=
T^{(-1/2)}_{\displaystyle(f\frac{d}{dx},a)}\;\circ\;{\cal L}-
{\cal L}\;\circ\;T^{(1/2)}_{\displaystyle(f\frac{d}{dx},a)}
\label{com}
\end {equation}\par

Let us give the explicit formula of this action.

\proclaim Proposition 1. The result of the action (\ref{com}) is an operator
of multiplication by the matrix:
\begin{equation}
\Bigl[ T_{\displaystyle(f\frac{d}{dx},a)},{\cal L}\Bigr]=\left(
\matrix{
\matrix{
fu'+2f'u-c_1f'''\hfill  \cr
+va'+c_2a''\hfill  \cr} & 
\matrix{
fv'+f'v-c_2f''\hfill \cr
+2c_3a'\hfill \cr}\hfill\cr\noalign{\bigskip}
\matrix{
fv'+f'v-c_2f'' \hfill  \cr
+2c_3a'\hfill  \cr} & 0\hfill\cr\noalign{\bigskip}
}\right)
\label{ad*}
\end{equation}\par
{\bf Proof}: straightforward.

\vskip 0,3cm

The following result clarifies the nature of the definition (\ref{com}).
It turns out that in the case of the Lie algebra {\goth \g}
the situation is analogue to those in the Virasoro case:
one obtains a generalization of the Kirillov-Segal result.

\proclaim Theorem 1. The action (\ref{com}) coincides with the coadjoint action of the 
Lie algebra {\goth \g}.
\par

We will prove this theorem in the next section.

\subsection{Coadjoint representation of the Lie algebra  {\Goth \g}}

Let us calculate the coadjoint action of the Lie algebra {\goth g}.

\vskip 0,3cm

{\bf Definition}.
Define the {\it regular part} of 
the dual space {\goth \g}$^*$ to the Lie algebra
{\goth \g} as follows (cf. \cite{Kir1}).
Put:
{\goth \g}$^*_{reg}=C^{\infty}(S^1)\oplus C^{\infty}(S^1)\oplus {\bf R}^3$
and fix the pairing 
$\langle\;,\;\rangle$: {\goth \g}$^*_{reg}\otimes${\goth \g} $\to{\bf R}$:
$$
\langle(u(x),\;v(x),\;{\bf c}),\;(f(x)\displaystyle\frac{d}{dx},\;a(x),\;{\bf \alpha})\rangle
=\displaystyle\int_{S^1}f(x)u(x)dx +\int_{S^1}a(x)v(x)dx + 
{\bf \alpha}\cdot{\bf c},
$$
where ${\bf c}=(c_1,c_2,c_3),\;{\bf \alpha}=(\alpha_1,\alpha_2,\alpha_3)\in {\bf R}^3$.

\vskip 0,3cm

\proclaim Proposition 2. The coadjoint action of {\goth \g} 
on the regular part of its dual space {\goth \g}$^*_{reg}$
is given by : 
\begin{equation}
ad^*_{\displaystyle(f\frac{d}{dx},a)}
\left(
\matrix{
u\hfill\cr\noalign{\smallskip}
v\hfill\cr\noalign{\smallskip}
{\bf c}\hfill\cr\noalign{\smallskip}}
\right)=
\left(
\matrix{
fu'+2f'u-c_1f'''+va'+c_2a''\hfill\cr\noalign{\smallskip}
fv'+f'v-c_2f''+2c_3a'\hfill\cr\noalign{\smallskip}
0\hfill\cr\noalign{\smallskip}}\right)
\label{coa}
\end{equation}
where ${\bf c}=(c_1,c_2,c_3)$ (the center of {\goth \g} acts trivially).\par

\vskip 0,3cm

{\bf Proof}. By definition of the coadjoint action, 
$\langle ad^*_{(f\frac{d}{dx},a)}(u,v,{\bf c}),\;((g\frac{d}{dx},b)\rangle=
-\langle(u,v,{\bf c}),\;[(f\frac{d}{dx},a),((g\frac{d}{dx},b)]\rangle$.
Integrate by part to obtain the result.

\vskip 0,3cm

The right hand side of the formula (\ref{coa}) 
coincides with the action (\ref{com}) of the Lie algebra 
$Vect (S^1)${\math \s}$ C^{\infty}(S^1)$ 
 on space of operators (\ref{ope}).

\vskip 0,3cm

Theorem 1 follows now from Proposition 1.

\vskip 0,3cm

{\bf Remark}. As a $\Vect(S^1)$-module, {\goth \g}$^*_{reg}$
is a deformation of the module
${\cal F}_{-2}\oplus {\cal F}_{-1}\oplus{\bf R}^3$
(and coincides with it if $c_1=c_2=0$).
Therefore, the dual space to the Lie algebra
has the following tensor sense: 
$$
u=u(x)(dx)^2, v=v(x)dx.
$$

The space of matrix Sturm-Liouville operators (\ref{ope})
gives a natural geometric realization of the dual space
to the Lie algebra {\goth g}.

\section{Generalized Neveu-Schwartz superalgebra}

We introduce here a Lie superalgebra which contains {\goth \g} 
as its even part.The relation between  {\goth \g} and this superalgebra
is the same as between the Virasoro algebra and the Neveu-Schwartz superalgebra.
We show that the differential operator (\ref{ope}) 
appears as a part of the coadjoint action of the constructed Lie
superalgebra.

We follow here the Kirillov method (see \cite{Kir2}) where the Sturm-Liouville operator
is realized as the even part of the coadjoint action of the Neveu-Schwartz superalgebra.

\subsection{Definition} 
 Consider the ${\bf Z}_2$-gradued vector space : 
 ${\cal S}={\cal S}_0\oplus{\cal S}_1$.
where 
${\cal S}_0= ${\goth \g} $=\Vect(S^1)\oplus C^{\infty}(S^1)\oplus{\bf R}^3$
 and ${\cal S}_1=C^{\infty}(S^1)\oplus C^{\infty}(S^1)$.
Define the structure of a Lie superalgebra on ${\cal S}$.

{\bf 1}. Define the action of the even part ${\cal S}_0$ on ${\cal S}_1$ by:
$$
\Bigl[(f(x)\frac{d}{dx},\;a(x)),\;(\phi(x),\;\alpha(x))\Bigr]:=
T^{(\frac{1}{2})}_{\displaystyle(f(x)\frac{d}{dx},\;a(x))}(\phi(x),\;\alpha(x))
$$
so that as a $\Vect(S^1)$-module
$$
{\cal S}_1={\cal F}_{\frac{1}{2}}\oplus{\cal F}_{-\frac{1}{2}}.
$$

{\bf 2}. 
The even part ${\cal S}_0$ acts on ${\cal S}_1$ according to (\ref{act'}).
Let us define the 
{\it anticommutator}
$[\;,\;]_+:{\cal S}_1\otimes{\cal S}_1\to{\cal S}_0$:
\begin{equation}
\Bigl[(\phi,\;\alpha),\;
(\psi,\;\beta)\Bigr]_+=
(\phi\psi\displaystyle\frac{d}{dx},\;\phi\beta+\alpha\psi,\;
{\bf \sigma_+})
\label{[]+}
\end{equation}
where ${\bf \sigma_+}=(\sigma_{+1},\sigma_{+2},\sigma_{+3})$ 
is the continuation of the cocycles (\ref{coc})
to the even part of ${\cal S}_0\subset{\cal S}$ defined by the formul{\ae}:
\begin{equation}
\matrix{
\sigma_{+1}((\phi,\alpha),(\psi,\beta))=
2\displaystyle\int_{S^1}\phi'(x)\psi'(x)dx\hfill\cr\noalign{\bigskip}
\sigma_{+2}((\phi,\alpha),(\psi,\beta))=
-2\displaystyle\int_{S^1}(\phi'(x)\beta(x)+\alpha(x)\psi'(x))dx\hfill\cr\noalign{\bigskip}
\sigma_{+3}((\phi,\alpha),(\psi,\beta))=
4\displaystyle\int_{S^1}\alpha(x)\beta(x)dx\hfill\cr\noalign{\bigskip}
}
\label{coc+}
\end{equation}

\proclaim Theorem 2.  ${\cal S}$ is a Lie superalgebra.\par

{\bf Proof}: One must verify the Jacobi identity :
\begin{equation}
 (-1)^{|X||Z|}[X,[Y,Z]]+(-1)^{|X||Y|}[Y,[Z,X]]
+(-1)^{|Y||Z|}[Z,[X,Y]=0,
\label{jac}     
\end{equation}
where
$|X|$ is the degree of $X$ ($|X|=0$ for $X\in {\cal S}_0$
and $|X|=1$ for $X\in {\cal S}_1$).

Let us prove (\ref{jac}) for  $ X,Y,Z\in{\cal S}_1$.
Take $X=(\phi,\alpha),\;Y=(\psi,\beta),\;Z=(\tau,\gamma)$,then:
$$
(a)\;\;\;\;\;\;\;\;\;\;\;\;\;\;\;\;\;\;\;\;\;\;\;\;
[(\phi,\alpha),[(\psi,\beta),(\tau,\gamma)]]=
-T^{1/2}_{[(\psi,\beta),(\tau,\gamma)]_+}(\phi,\alpha).
$$
The expression $[(\psi,\beta),(\tau,\gamma)]_+$
is given by (\ref{jac}) one gets:
$T^{1/2}_{[(\psi,\beta),(\tau,\gamma)]_+}(\phi,\alpha)=
T^{1/2}_{(\psi\tau,\psi\gamma+\beta\tau)}(\phi,\alpha).$
According to (\ref{act'}),
$$
T^{1/2}_{(\psi\tau,\psi\gamma+\beta\tau)}(\phi,\alpha)
=(L_{\psi\tau}^{1/2}(\phi),\;
L_{\psi\tau}^{-1/2}(\alpha) -\frac{1}{2}(\psi\gamma+\beta\tau)'{\phi}).
$$
where $L_{\psi\tau}^{1/2}(\phi)=\psi\tau\phi'-\frac{1}{2}(\psi'\tau+\psi\tau')\phi$
and $ L_{\psi\tau}^{-1/2}(\alpha) -\frac{1}{2}(\psi\gamma+\beta\tau)'{\phi}= 
\psi\tau\alpha'+\frac{1}{2}(\psi\tau)'\alpha-\frac{1}{2}(\psi\gamma)'\phi-
\frac{1}{2}(\beta\tau)'\phi.$

In the same way we obtain:
$$
(b)\;\;\;\;\;\;\;\;\;\;\;\;[(\psi,\beta),[(\tau,\gamma),(\phi,\alpha)]]
=(L_{\phi\tau}^{1/2}(\psi),\;
L_{\phi\tau}^{-1/2}(\beta) -\frac{1}{2}(\tau\alpha+\phi\gamma)'{\psi})
$$
where $L_{\phi\tau}^{1/2}(\psi)=\phi\tau\psi'-\frac{1}{2}(\phi'\tau+\phi\tau')\psi$
and $L_{\phi\tau}^{-1/2}(\beta) -\frac{1}{2}(\tau\alpha+\phi\gamma)'{\psi}=
\phi\tau\beta'+\frac{1}{2}(\phi\tau)'\beta-\frac{1}{2}(\tau\alpha)'\psi-\frac{1}{2}(\gamma\phi)'\psi.$

For the last term one has:
$$
(c)\;\;\;\;\;\;\;\;\;\;\;\;[(\tau,\gamma),[(\phi,\alpha),(\psi,\beta)]]
=(L_{\phi\psi}^{1/2}(\tau),\;
L_{\phi\psi}^{-1/2}(\gamma) -\frac{1}{2}(\phi\beta+\psi\alpha)'{\tau})
$$
where $L_{\phi\psi}^{1/2}(\tau)=\phi\psi\tau'-1/2(\phi'\psi+\phi\psi')\tau$
and $L_{\phi\psi}^{-1/2}(\gamma) -{1/2}(\phi\beta+\psi\alpha)'{\tau}=
\phi\psi\gamma'+1/2(\phi\psi)'\gamma-1/2(\phi\beta)'\psi-1/2(\alpha\psi)'\tau.$

Taking the sum $(a)+(b)+(c)$ one obtains zero.

The proof of the Jacobi identity for the other cases is analogue.

\vskip 0,3cm

Theorem 2 is proven.

\proclaim Proposition 3. The coadjoint action of  ${\cal S}$
is given by the formula :
$$
ad^*_{\left(
\matrix{
f\frac{d}{dx}\hfill\cr
a\hfill\cr
\phi (dx)^{-\frac{1}{2}}\hfill\cr
\alpha (dx)^{\frac{1}{2}}\hfill\cr}
\right)}
\left(\matrix{
u\hfill\cr\noalign{\bigskip}
v\hfill\cr\noalign{\bigskip}
{\bf c}\hfill\cr\noalign{\smallskip}
\psi \hfill\cr\noalign{\bigskip}
\beta \hfill\cr\noalign{\bigskip}
}
\right)
=\left(
\matrix{
\matrix{
L^{(-2)}_f(u)+va'+c_2a''-c_1f'''\hfill\cr
+\frac{1}{2}\psi'\phi+\frac{3}{2}\psi\phi'-\frac{1}{2}\beta'
\alpha+\frac{1}{2}\beta\alpha'\hfill\cr
}\hfill\cr\noalign{\smallskip}
\matrix{
L^{(-1)}_f(v)+2c_3a'-c_2f''\hfill\cr
+\frac{1}{2}\beta'\phi+\frac{1}{2}\beta\phi'\hfill\cr
}\hfill\cr\noalign{\smallskip}
0\hfill\cr\noalign{\smallskip}
\matrix{
L^{(-3/2)}_f(\psi)+\frac{1}{2}a'\beta\hfill\cr
-2c_1\phi''+u\phi+v\alpha+2c_2\alpha'\hfill\cr
}\hfill\cr\noalign{\smallskip}
\matrix{
L^{(-1/2)}_f(\beta) \hfill\cr
-2c_2\phi'+v\phi+4c_3\alpha\hfill\cr
}\hfill\cr\noalign{\smallskip}
}
\right)
$$
where ${\bf c}=(c_1,c_2,c_3)$ (as usual, the center acts trivially).\par

{\bf Proof}: difect calculation using the definition of the superalgebra $S$.

\vskip 0,3cm

In particular, one obtains:
\proclaim  Corollary. 
$$
ad^*_{\left(\matrix{
0\hfill\cr
0\hfill\cr
\phi(dx)^{-\frac{1}{2}}\hfill\cr
\alpha(dx)^{\frac{1}{2}}\hfill\cr
}
\right)}
\left(
\matrix{
u\hfill\cr
v\hfill\cr
{\bf c}\hfill\cr
0\hfill\cr
0\hfill\cr
}\right)
=\left(
\matrix{
0\hfill\cr
0\hfill\cr
0\hfill\cr
-2c_1\phi''+u\phi+v\alpha+2c_2\alpha'\hfill\cr
-2c_2\phi'+v\phi+4c_3\alpha \hfill\cr
} 
\right)
$$\par
This corollary gives the matrix operator (\ref{ope}) defined in Section 2.

\vskip 0,3cm

The Lie superalgebra ${\cal S}$ seems to be an interesting generalization
of the Neveu-Shwartz superalgebra. It would be interesting to obtain some
information about its representations, coadjoint orbits, corresponding
integrable systems etc.

\vskip 1cm

%%%%%%%%%%%%%%%%%%%%%%%%%%%%%%%%%%%%%%%%%%%%%%%%%%%%%%%%%%%%%%%%%%%%%%%%%%%%%%
%%%%%%%%%%%%%%%%%%%%%%%%%%%%%%%%%%%%%%%%%%%%%%%%%%%%%%%%%%%%%%%%%%%%%%%%%%%%%%

\end{document}